\documentclass[
    reprint,
    superscriptaddress,
    amsmath,amssymb,
    aps,
    prx,
    longbibliography,
]{revtex4-1}

\usepackage{graphicx}
\usepackage{amsmath}
\usepackage{amssymb}
\usepackage[colorlinks=true,urlcolor=blue,linkcolor=blue,citecolor=blue,breaklinks]{hyperref}
\usepackage[margin=0.75in]{geometry}

\begin{document}

\title{Simulation of quantum many-body dynamics with\\ Tensor Processing Units:  Floquet prethermalization}

\author{Alan Morningstar}
\affiliation{Department of Physics, Princeton University, Princeton, NJ 08544, USA}
\affiliation{Sandbox@Alphabet, Mountain View, CA 94043, USA}

\author{Markus Hauru}
\affiliation{Sandbox@Alphabet, Mountain View, CA 94043, USA}

\author{Jackson Beall}
\affiliation{Sandbox@Alphabet, Mountain View, CA 94043, USA}

\author{Martin Ganahl}
\affiliation{Sandbox@Alphabet, Mountain View, CA 94043, USA}

\author{Adam G. M. Lewis}
\affiliation{Sandbox@Alphabet, Mountain View, CA 94043, USA}

\author{Vedika Khemani}
\affiliation{Department of Physics, Stanford University, Stanford, CA 94305, USA}

\author{Guifre Vidal}
\affiliation{Sandbox@Alphabet, Mountain View, CA 94043, USA}

\date{\today}

\begin{abstract}

Tensor Processing Units (TPUs) are specialized hardware accelerators developed by Google to support large-scale machine-learning tasks, but they can also be leveraged to accelerate and scale other linear-algebra-intensive computations. In this paper we demonstrate the usage of TPUs for massively parallel, classical simulations of quantum many-body dynamics on long timescales. We apply our methods to study the phenomenon of \textit{Floquet prethermalization}, i.e., exponentially slow heating in quantum spin chains subject to high-frequency periodic driving. We simulate the dynamics of $L=34$ qubits for over $10^5$ Floquet periods, corresponding to circuits with $4\times10^6$ nearest-neighbor two-qubit gates. The circuits simulated have no additional symmetries and represent a pure-state evolution in the full $2^L$-dimensional Hilbert space. This is achieved by distributing the computation over 128 TPU cores. On that size TPU cluster, we find speedups in wall-clock runtime of $230\times$ and $15\times$ when compared to reference CPU and single-GPU simulations, respectively, for shorter-time 30-qubit simulations that can be handled by all three platforms. We study the computational cost of the simulations, as a function of both the number of qubits and the number of TPU cores used, up to our maximum capacity of $L=40$ qubits, which requires a ``full pod" of 2048 TPU cores with tens of terabytes of memory in total. For these simulations, an 8-TPU-core machine is comparable to a single A100 GPU, and thus the full TPU pod is comparable to a machine with hundreds of top-of-the-line GPUs. However, the TPU pod is more energy and cost efficient, and readily accessible (via Google Cloud), unlike such large many-GPU configurations. We also study the accumulation of numerical error as a function of circuit depth in very deep circuits. Our work demonstrates that TPUs can offer significant advantages for state-of-the-art simulations of quantum many-body dynamics.

\end{abstract}

\maketitle

\section{Introduction}

The study of out-of-equilibrium quantum dynamics is an exciting research frontier. Recent years have seen many important developments on foundational questions ranging from the onset of chaos and thermalization in isolated quantum systems~\cite{Polkovnikov-Vengalattore2011,DAlessio-Rigol2016}, to mechanisms for understanding the breakdown of thermalization such as many-body localization~\cite{Nandkishore-Huse2015,Abanin-Serbyn2019,Alet-Laflorencie2018} or quantum scarring~\cite{Moudgalya-Regnault2021, DimaScarsReview}, to the discovery of novel non-equilibrium phases of matter~\cite{HarperReview2020}. Research in this area involves the study of the \emph{out-of-equilibrium dynamics} of \emph{highly-excited} many-body quantum systems, which lies outside the purview of most traditional frameworks and techniques of many-body quantum theory, thereby necessitating the development of a new conceptual and technical toolkit. 

Simulating the dynamics of a general quantum many-body system on a classical computer is prohibitively costly due to the exponential scaling of the dimension of the quantum state space with system size. This reality, along with the fact that quantum mechanical systems are of both fundamental interest and practical importance, was the original motivation to begin the longstanding investigation and development of computers that operate quantum mechanically~\cite{Feynman1982}. While significant progress has been made in developing quantum simulators across a variety of platforms ~\cite{Georgescu-Nori2014,Ladd-OBrien2010,Bloch-Nascimbene2012,Blatt-Roos2012,Monroe-Yao2021,AspuruGuzik-Walther2012,Houck2012,Preskill2018,Altman-Zwierlein2021,Alexeev-Thompson2021,Awschalom-Zhang2021}, much research in quantum dynamics still relies heavily on classical numerical simulations. 

In parallel to the development of quantum hardware, the rise in prominence of computational science and engineering---and in particular, data-intensive machine learning---has led to advances in classical computing, namely, the integration of specialized \textit{hardware accelerators} into supercomputing systems, to meet these growing needs. The most common example of a hardware accelerator is the Graphics Processing Unit (GPU), which is a widely available type of general-purpose processor capable of performing many arithmetic operations in parallel. Another is Google's development of the Tensor Processing Unit (TPU), and the JAX and TensorFlow software libraries, as a means of accelerating and scaling the computations needed to train and evaluate large neural networks with accessible tools~\cite{TPUinfo, jouppi2017datacenter, jax, Frostig-Leary2018, Abadi-Zheng2016}. Unlike GPUs, TPUs are not general-purpose processors; they are specialized application-specific integrated circuits (ASICs) for deep learning workloads that are heavy on matrix multiplication and vector operations.

Given these advances in classical computing capabilities, it is natural to explore the application of TPUs to other challenging computational tasks~\cite{Belletti-Anderson2020,Wang-Anderson2021,Pan-Mishra2021,Lu-Ma2020,TPUFFT1,TPUFFT2,huot2019highresolution}. Since TPUs were specifically designed for large-scale linear algebra, the classical simulation of quantum systems is a particularly suitable application for them. More specifically, the strength of using TPUs for simulating the dynamics of quantum systems comes from two contributions: (1) TPU cores have hardwired matrix multiplication units that allow them to contract arrays (with some constraints on their shapes) faster and with higher energy efficiency than on contemporary CPUs and GPUs~\cite{jouppi2017datacenter}. (2) Arrays can be distributed over thousands of directly linked TPU cores, and operated on easily within the single-program-multiple-data (SPMD) programming paradigm~\footnote{Single program, multiple data (SPMD): A parallel programming paradigm whereby a single set of instructions (program) is evaluated simultaneously on multiple distinct inputs (data) hosted on a set of parallel processes.}. Data transfer between the cores is carried out via fast inter-core interconnects (ICIs), and this makes multi-TPU configurations extraordinarily performant at scale. These ingredients naturally allow for the accelerated application of local operators to wavefunctions that are distributed over many TPU cores, and other useful operations. Similar efforts have been made to leverage GPUs to accelerate the classical simulation of quantum circuits~\cite{Li-Krishnamoorthy2020,Vincent-Dhand2021,Efthymiou-Carrazza2020,SuzukiFujii2020,Jones2019,Luo-Wang2020,Isakov-Boixo2021,Pan-Zhang2021,HybridQ,Kelly2018,Huang2020,quimb,Villalonga-Mandra2020}, also resulting in large gains over CPU simulations. TPUs exist because they are more performant and more energy efficient (and thus cost effective) than GPUs for many large-scale machine-learning tasks. It is therefore promising to begin to explore their application to quantum simulation.

In this work, which is the first of a set of related papers~\cite{tpu_algebra,tpu_circuit,tpu_qphys,tpu_qhardware,tpu_qchem,tpu_spinLED,tpu_DMRG,tpu_Z2field}, we demonstrate the application of TPUs to accelerate and scale the classical simulation of quantum many-body dynamics. Specifically, we study dynamics in Floquet, i.e., time-periodic, quantum circuits as a technically illustrative and physically motivated application. Periodic driving has become a central workhorse of quantum simulation, both as a tool for ``Floquet engineering" interesting band structures and Hamiltonians~\cite{Okay-Kitamura2019}, and as a means to realize novel intrinsically non-equilibrium phases of matter such as time-crystals~\cite{HarperReview2020, Bukov_2015, Oka_2009, Rudner2013,AFAI,Khemani2016, Else-Nayak2016, Po2016}. Indeed, observing such novel phases of driven systems was recently identified~\cite{Ippoliti2021} and experimentally demonstrated~\cite{GoogleDTC_2021} as an exceptionally well-suited near-term application of digital quantum simulators for many-body physics. Thus pushing the boundaries of the classical simulation of Floquet systems is not only motivated by a fundamental theoretical interest in quantum dynamics, but also by a need to benchmark quantum simulators, and to raise the bar for evaluating quantum advantage.

We focus on the phenomenon of \textit{Floquet prethermalization} for our demonstration. Periodically driving an interacting many-body system generically leads to thermalization to a featureless infinite-temperature state~\cite{DAlessio-Rigol2014, Lazarides-Moessner2014}; the time dependence of the Hamiltonian breaks the usual conservation of energy, and infinite temperature is the appropriate maximum-entropy equilibrium in the absence of any additional conservation laws. The heating to infinite temperature can be averted in the presence of disorder induced many-body localization~\cite{Ponte_2015, Lazarides_2015}. Alternatively, at sufficiently high driving frequencies $\omega$, the heating can be slowed down to a rate that is exponentially small in $\omega$ corresponding to a heating time $t_h \sim e^{\omega/\omega_0}$~\cite{Abanin-Huveneers2015,Mori-Saito2016,Kuwahara-Saito2016,Abanin-Huveneers2017,Abanin-Huveneers2017b,Else-Nayak2017, Luitz-Khemani2020, RubioAbadal-Bloch2020, Peng-Cappellaro2021}. 
This high-frequency prethermal regime with exponentially slow heating is challenging to observe in finite-size numerics~\cite{Machado-Yao2019,Luitz-Khemani2020}: One needs driving frequencies that are large compared to the local energy scales $\sim J$ in the problem, but small compared to the extensive many-body bandwidth $\sim J L$ (in 1D systems of length $L$), which does not present a large dynamic range in systems of size amenable to numerical simulation. Furthermore, exponentially long time evolutions are needed, and the full $2^L$-dimensional wavefunction must be used (rather than a truncated tensor network representation) because of the extensive amount of entanglement generated by the dynamics. For these reasons, the Floquet prethermalization problem is a well-motivated choice for demonstrating the technical advantages of using TPUs to accelerate and scale up classical simulations of the quantum dynamics of Floquet circuits.

The rest of this work is structured as follows: In Sec.~\ref{sec:algorithms} we provide an overview of the conceptual principles of quantum simulation algorithms for TPUs. Then in Sec.~\ref{sec:model} we describe the model we use and provide a brief review of Floquet prethermalization. Following this, in Sec.~\ref{sec:sims_results} we present the physics-centric results of our numerical simulations. Then, in Sec.~\ref{sec:cost_accuracy} we move on to a study of the computational cost of, and accumulation of numerical errors in, the simulations done in this work. Finally, we summarize our work and discuss potential future applications of TPU simulations to quantum dynamics research in Sec.~\ref{sec:summary}.

\section{Principles of TPU algorithms \label{sec:algorithms}}
In later sections we study the long-time dynamics of periodically driven quantum spin chains using TPU simulations, so here we first provide a high-level overview of TPUs and the principles of using TPUs for these simulations.

The computing platform we work with is a cluster of $2^{N_g}$ connected TPU v3 cores, with $N_g\in[3,11]$ (number of cores $\in [8,2048]$). The connectivity of the cores is toroidal, and each core has 16 GB of high-bandwidth memory. A full 2048-core v3 ``pod" therefore has 32 TB of memory. Since simulating a larger number of qubits $L$ requires exponentially more memory (in $L$), we adjust $N_g$ based on the total memory needed and/or the total number of TPUs we want working simultaneously. Generally, using the highest fraction of memory possible (using the smallest cluster possible) results in the highest efficiency. On TPUs, matrix multiplication occurs natively in half-precision, but single and double precision can be emulated, and all of our simulations in this work are done in single precision.

\begin{figure}
    \centering
    \includegraphics[width=1\linewidth]{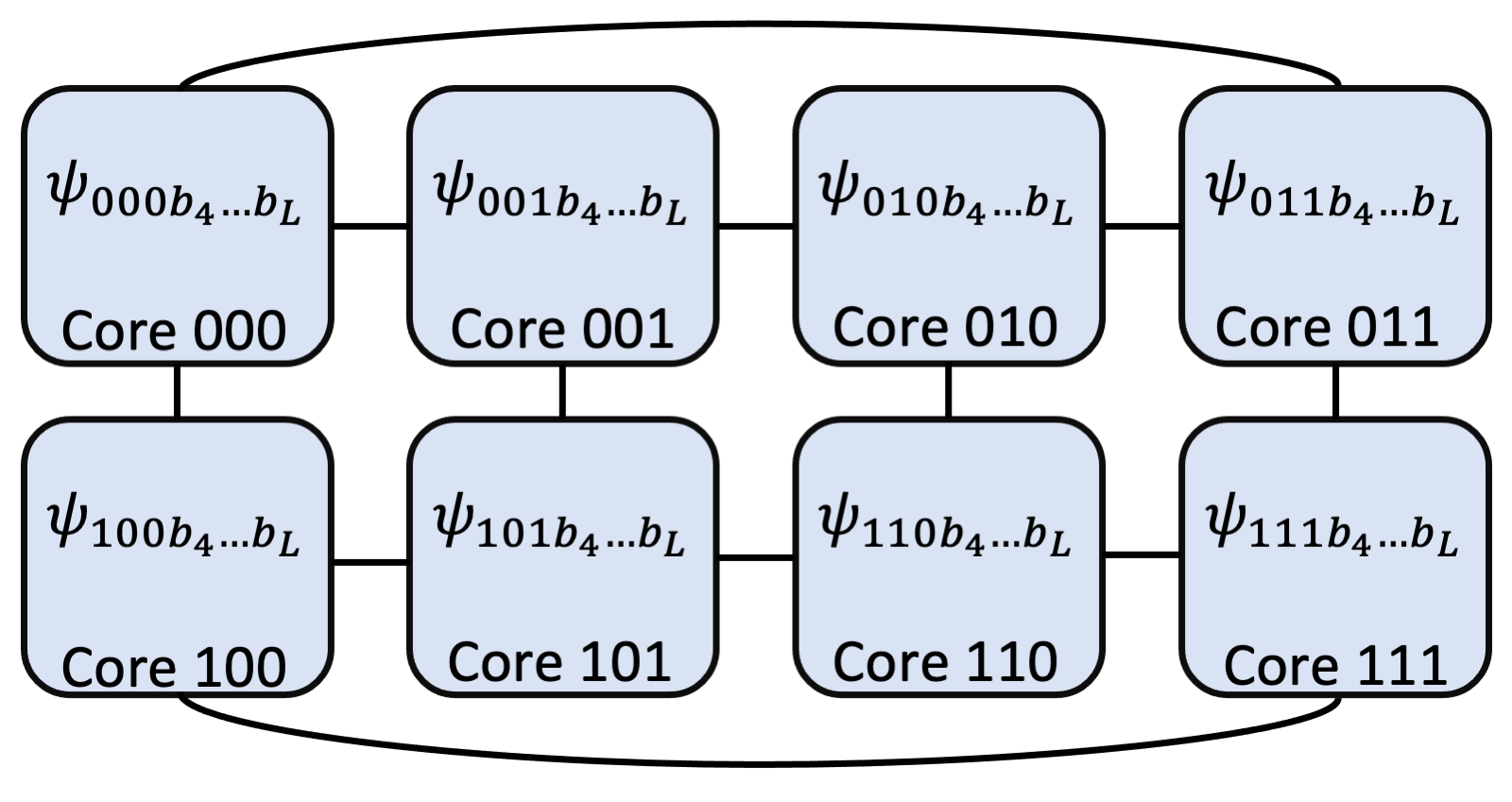}
    \caption{Distribution of the many-body wavefunction over TPU cores. There are eight TPU cores depicted ($N_g=3$), labelled by the binary strings $000,001,...,111$. On each core, a branch of the full wavefunction is stored, where the state of the global qubits is fixed according to the TPU core. Lines between TPU cores indicate the toroidal pattern of intercore connectivity.}
    \label{fig:distributed_wavefunction}
\end{figure}

For concreteness, we will consider the example of $N_g=3$ (eight TPU cores). Thus the cores can be labelled by three-bit binary numbers $c_1c_2c_3\in\{000,001, ..., 111\}$. The central object relevant to our simulations is the full wavefunction of $L$ qubits, which is distributed over all of the cores in the following way: The wavefunction is the array of amplitudes $\psi_{b_1...b_L} = \langle b_1...b_L| \psi \rangle$, where $b_1...b_L$ is a bit string that labels the computational basis states. The subset of these amplitudes that we store on a particular core is then simply the subset whose first three qubits, termed the \textit{global qubits}, match the core's three-bit label ($c_1c_2c_3=b_1b_2b_3$), i.e., each core stores its own branch of the wavefunction where the global qubits are in a definite computational state. For example, we store the branch of the wavefunction where the three global qubits are in state $|101\rangle$ on the TPU core labelled by $101$ (the sixth core), which corresponds to the array $\psi_{101b_4...b_L}$ existing locally on that core. In this way, each core only stores an array of size $2^{L-3}$ (or, more generally, $2^{L-N_g}$) locally. For this reason, we refer to the $L-N_g$ final qubits as \textit{local qubits}. Since $2^{29}$ single-precision complex floats take up 4.3 GB of memory, the 16 GB of memory per core is enough to store three separate wavefunctions with $L-N_g=29$ local qubits. The maximum total number of qubits we can do simulations with is therefore $L=40$, because the largest TPU v3 cluster corresponds to $N_g=11$ global qubits (2048 cores). See Fig.~\ref{fig:distributed_wavefunction} for a depiction of the distributed wavefunction.

We use the JAX software library~\cite{jax} to write SPMD-style code to execute our simulations on the TPU clusters. Recall that SPMD (single program, multiple data) means that the same set of instructions is run on each TPU core, but each core hosts a different set of data to input into those instructions. Notably, the ``single program" can include intercore communication as long as it is symmetric under permuting the cores. This is the type of parallelism that is available in JAX (see the documentation for available functionality, tutorials, etc.~\cite{jax_docs}), and it can be used to do all of the necessary operations, ex: prepare initial states, apply local operators, compute inner products, etc. TPUs have a native matrix multiplication unit and data storage pattern that works optimally for arrays whose shape satisfies certain criteria: the basic rules of thumb are that the final two dimensions of an array must have sizes that are multiples of 8 and 128, respectively, and contractions must occur on indices whose size is a multiple of 128. Note that $8=2^3$ (three qubits) and $128=2^7$ (seven qubits), so these are practical constraints about how many qubits must be grouped together in a wavefunction array and during operations like applying local operators to wavefunctions. But here we will ignore that complication for the sake of clarity. Below we illustrate the principles of TPU algorithms for quantum simulation via two simple examples.

Let's first consider an inner product between two distributed wavefunctions $\psi$ and $\phi$. This can be written as (recall $N_g=3$ here)
\begin{eqnarray}
    \langle\psi|\phi\rangle=\sum_{c_1c_2c_3}\sum_{b_4...b_L}\psi^*_{c1c2c3b_4...b_L}\phi_{c1c2c3b_4...b_L}.\label{eq:inner_product}
\end{eqnarray}
When evaluating this, the inner sum over local qubits occurs first locally on each core, i.e., an inner product (which is a ``single program") of the branches of the wavefunctions on each TPU core (the ``multiple data") is evaluated resulting in a single number on each core that can be indexed by the label of that core $c_1c_2c_3$ (another ``multiple data" form). The outer sum over the state of the global qubits in Eq.~\ref{eq:inner_product} is then evaluated simply by collecting all of these numbers from all of the individual cores and summing them. Each core does this identically (another ``single program" that runs the same on each core). This can be done using JAX's \texttt{psum} operation~\cite{jax_docs}, which is one of the few operations that encodes SPMD-type intercore communication.

As a second and final example, we'll consider applying a local two-qubit gate. Again, in practice, to get optimal performance from the TPUs we must fuse gates into operators acting on certain groups of seven or more qubits, but we ignore that detail here for conceptual clarity. In the case that the gate acts only on local qubits, the action of applying the gate (single program) can be performed independently on each branch of the wavefunction (multiple data) that is local to each core. For example, applying a gate to qubits 7 and 8 results in the updated local (to each core labelled by $c_1c_2c_3$) branches being
\begin{eqnarray}
    (U\psi)_{c_1c_2c_3b_4...b_L} = \sum_{\alpha\beta} U_{b_7b_8\alpha\beta} \psi_{c_1c_2c_3b_4...\alpha\beta...b_L}. 
\end{eqnarray}
Notice that the same sum over $\alpha$ and $\beta$ can occur independently on each core with the data from the gate and the locally stored branch of the wavefunction. The result is an updated wavefunction that has the same distribution pattern as the original one. In the case where the gate must be applied to some of the global qubits, we use JAX's \texttt{pswapaxes}~\cite{jax_docs} (which allows intercore communication) to effectively swap the global qubits that label the TPU cores with the local qubits that do not. This reduces this case to the previous one of applying a gate to local qubits.

Much of the effort in writing these algorithms goes into complying with the aforementioned constraints implied by requiring optimal performance from the TPU matrix multiplication unit. More discussion of this can be found in subsequent works~\cite{tpu_qphys,tpu_algebra,tpu_circuit}, but what has been provided here is all that is needed for the simulations in the following sections.

\section{Model and review \label{sec:model}}
In this section we provide both a concrete model that we will work with for the rest of the paper, and a review of the relevant physics of Floquet prethermalization.

\subsection{Floquet circuit model}
\begin{figure}
    \centering
    \includegraphics[width=1\linewidth]{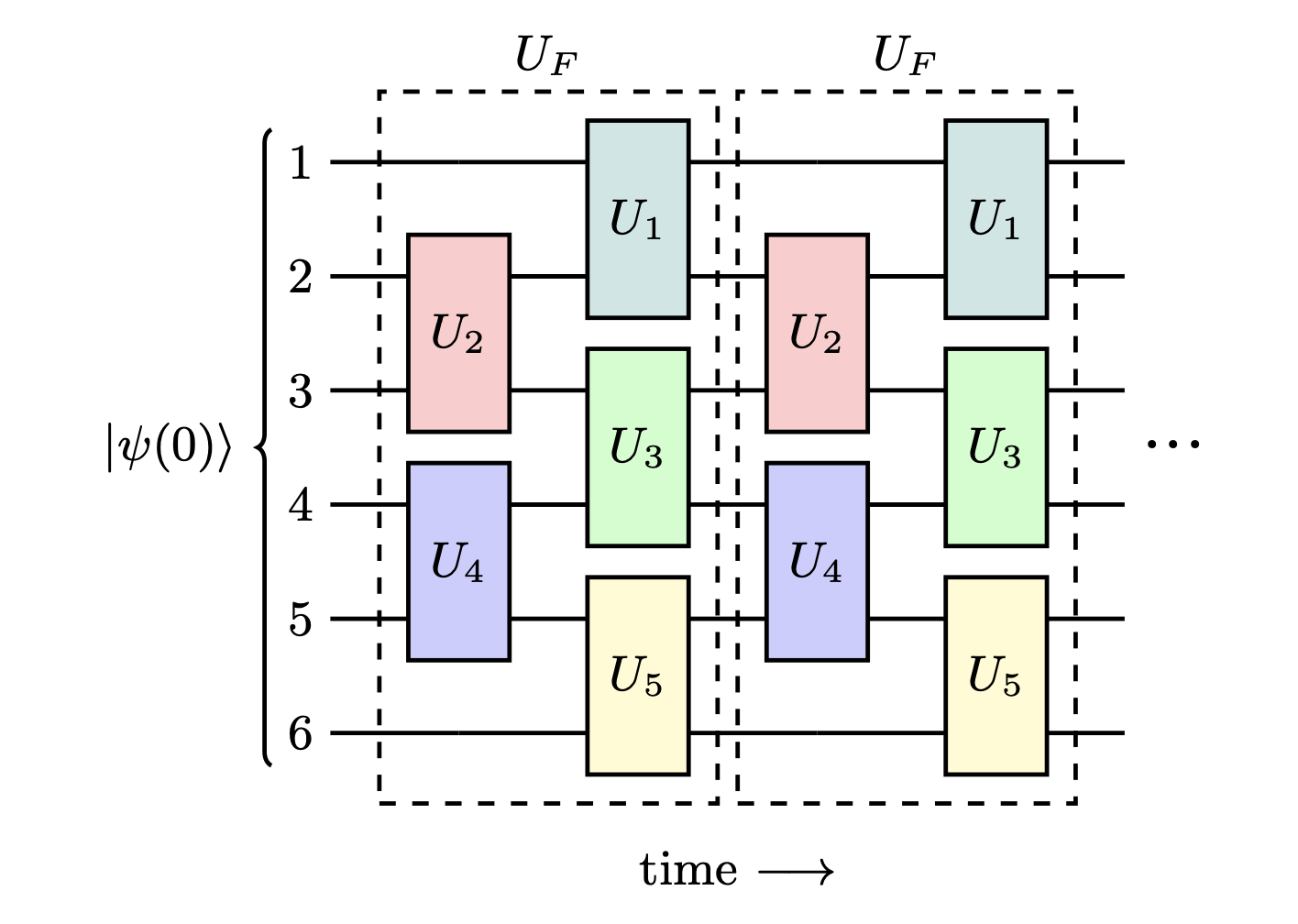}
    \caption{The time-periodic quantum circuit generating the dynamics of our model system, illustrated on $L=6$ qubits. Qubits are numbered top to bottom, and time runs left to right. The system is prepared in the state $|\psi(0)\rangle$ described later in the main text (Eq.~\ref{eq:init_state}). Each gate $U_k$ takes time $T/2$ to be executed, during which its two qubits are evolving under an associated two-qubit Hamiltonian $2H_k$ (Eq.~\ref{eq:gate}). All gates in the same layer are applied simultaneously, so the two-layer repeating unit (the Floquet unitary $U_F$) lasts for time $T$.}
    \label{fig:circuit}
\end{figure}
Our model system consists of $L$ spin-1/2 (qubit) degrees of freedom arranged in a one-dimensional array with open boundaries. The stroboscopic time evolution of the state of the system is generated by repeatedly applying a Floquet unitary $U_F$, which is the unitary time-evolution operator over one period, i.e., 
\begin{eqnarray}
    |\psi(nT)\rangle = U_F^n |\psi(0)\rangle,
\end{eqnarray}
where $T$ is the physical time duration of $U_F$ (the Floquet period), and $n\in \mathbb{N}$. The frequency is $\omega = 2 \pi / T$. 

We choose the Floquet unitary to be a two-layer quantum circuit of close-packed two-qubit unitary gates $U_k$ acting on qubits $k$ and $k + 1$ (see Fig.~\ref{fig:circuit}). The gates alternate between acting on even and odd bonds in each layer. 
For even $L$, this takes the form
\begin{equation}
    U_F = \prod_{k=1}^{L/2} U_{2k-1} \prod_{k=2}^{L/2} U_{2k-2}. 
\end{equation}
Each gate $U_k$ enacts Hamiltonian time evolution on its two qubits under the Hamiltonian $2H_k$ for time $T/2$,  
\begin{eqnarray}
    U_k = \exp(-i H_k T),\label{eq:gate}
\end{eqnarray}
where we work in units where $\hbar = 1$. 

Since our circuit is time-periodic (Floquet), with period $T$, and the gates are formulated in terms of two-qubit Hamiltonians $H_k$, we can also view the system as evolving under the following time-dependent local Hamiltonian $H(t)$, with $H(t) = H(t + T)$:
\begin{eqnarray}
    &H(t)  = \begin{cases} 
    2H_e, \quad n \leq \frac{t}{T} < n+\frac{1}{2}\\
    2H_o,  \quad n+\frac{1}{2} \leq \frac{t}{T} < n+1
    \end{cases}&\nonumber\\
    &H_e \equiv \sum_{k=2}^{L/2} H_{2k-2}
    \hspace{0.25in}H_o \equiv \sum_{k=1}^{L/2} H_{2k-1}.&
    \label{eq:Ht}
\end{eqnarray}
During the first half of each period, $H(t) = 2 H_e$, and during the second half, $H(t) = 2 H_o$, where $H_e$ and $H_o$ contain the even and odd-bond terms of the time-averaged Hamiltonian operator
\begin{eqnarray}
    \bar{H} = \sum_{k=1}^{L-1} H_k. \label{eq:hamiltonian}
\end{eqnarray} 

The $L-1$ independent two-qubit Hamiltonians $H_k$ are each chosen by first sampling a random Hermitian $4\times4$ matrix from the Gaussian unitary ensemble (GUE), then shifting and scaling that matrix to be traceless with Frobenius norm $J=\sqrt{2}$, which sets the microscopic energy scale $J$ and the units of energy and time that we use. Thus our model is actually an ensemble of random Floquet circuits that has a single tuning parameter $T$ (or $\omega$). Random Floquet circuits have been used as models of both quantum chaos~\cite{Chan-Chalker2018} and many-body localization (MBL)~\cite{Sunderhauf-Cirac2018}, and we emphasize that the randomness in our model does not induce any form of MBL. In order to avoid unnecessary ensemble averaging, we will only use a single sample from the ensemble in this paper, because a single sample is already a good representative of a generic, interacting, driven system. This means that we sample a long list of two-qubit Hamiltonians once, and then use the first $L-1$ terms from that list for an $L$-qubit system. This is a desirable property when assessing convergence in $L$, while avoiding the need to average over samples~\footnote{To be sure that we have not chosen an atypical sample with, for example, very weak interactions or very strong single-qubit potentials anywhere, we have split each $H_k$ into its component noninteracting and interacting pieces and checked that the norms of these pieces are $\lesssim 1$ and $\sim 1$, respectively.}.

We choose to use this random Floquet circuit model because circuits are computationally convenient to simulate compared to continuous-time Hamiltonian evolution that cannot be written as a circuit, where an artificial discretization scheme is needed. Circuits are also the natural mode of operation of digital quantum simulators. Note that this circuit model can also be viewed as a trotterization~\cite{Lloyd1996,Suzuki1991,Vidal2004,Heyl-Zoller2019} of  evolution under the time-averaged (and time-independent) Hamiltonian $\bar{H}$. However, here we study the full dynamics of $U_F$ even outside of the regime of very small $T$ (large $\omega$) where the trotterization approximates evolution due to $\bar{H}$. Our choice of random $H_k$ is so as to study a minimal model with spatial locality and periodic driving, but no other physical features (such as symmetries, topology, etc.), in order to capture general universal features of high-frequency Floquet prethermalization.

\subsection{Floquet prethermalization}
As mentioned earlier, for any (finite) drive frequency $\omega$, if $L$ is large enough, a generic driven many-body system thermalizes to a featureless ``infinite temperature" thermal state~\cite{DAlessio-Rigol2014,Lazarides-Moessner2014}. This means that the reduced density matrix of any small subsystem will be proportional to the identity. However, if  $\omega \gg J$, i.e., the drive frequency is much larger than the local energy scale of the system, then the transitions that allow the system to exchange energy with the drive will involve the rearrangement of many local degrees of freedom; an estimate for how many is $m \sim \omega / J$ because each local transition is on an energy scale $J$ and the drive couples to energy $\omega$. In that case, the effective rates for these processes will be exponentially suppressed in $m$ because there are $\sim m$ off-resonant virtual states involved, and thus the thermalization rate (time) will be exponentially small (large) in $\omega/J$~\cite{Abanin-Huveneers2015,Mori-Saito2016,Kuwahara-Saito2016,Abanin-Huveneers2017,Abanin-Huveneers2017}.

More precisely: The Floquet unitary of our model is given by $U_F = \exp(-i H_o T)\exp(-i H_e T)$, but one can always define a (nonunique) Floquet Hamiltonian $H_F$ such that $U_F \equiv \exp(-i H_F T)$, i.e., the stroboscopic dynamics are identical to evolution generated by a time-independent Hamiltonian $H_F$. Since in the large-$L$ limit the system always thermalizes to an infinite-temperature equilibrium, the simultaneous eigenstates of $U_F$ and $H_F$ are all infinite-temperature states, and thus $H_F$ is a highly nonlocal, unphysical Hamiltonian. 
However, if $\omega \gg J$, then there is a \textit{local} Hamiltonian $\tilde{H}_F$ such that $||U_F - \exp(-i\tilde{H}_F T)||$ is exponentially small in $\omega/J$, and the dynamics of the system are well-approximated by the time-independent Hamiltonian $\tilde{H}_F$ out to times exponentially long in $\omega/J$. This is one of the main results of several mathematical works on this topic~\cite{Kuwahara-Saito2016, Mori-Saito2016, Abanin-Huveneers2017, Abanin-Huveneers2017b}.

In order to approximate $\tilde{H}_F$, one can employ a high-frequency Baker-Campbell Hausdorff (BCH) expansion to get
\begin{eqnarray}
    \tilde{H}_F = H_o + H_e - \frac{iT}{2} [H_o, H_e] + O(T^2 H^3),\label{eq:BCH}
\end{eqnarray}
where $H_o$ and $H_e$ are the odd and even-$k$ terms in Eq.~\ref{eq:Ht}. This is an asymptotic series which eventually diverges, but it appears to be converging up to a finite order $m_\text{opt}\sim \omega/J$, which defines an effective Hamiltonian, $\tilde{H}_F$, obtained from truncating the series. The zeroth order term in $\tilde{H}_F$ is the time-averaged Hamiltonian $\bar{H} = H_o + H_e$, which provides a useful approximate notion of energy in our system, which we will use throughout the rest of this work. The exponentially long thermalization time will be detected by monitoring the evolution of $\langle \bar{H} \rangle$ over time, which is $\langle \tilde{H}_F \rangle$ up to relative corrections of $O(JT)$ [absolute corrections of $O(JTH)$]. Note that once the frequency becomes comparable to the (extensive) many-body bandwidth, a high-frequency expansion can be convergent in which case one obtains a sensible local $\tilde{H}_F$ governing the dynamics for all time; however, this is not expected to happen for finite frequencies for systems in the large $L$ limit. 

This discussion illustrates why it is numerically challenging to observe prethermalization:  simulations need to reach long enough times to study the exponentially slow processes involved, and large enough system sizes to avoid finite-size effects. Thus this problem provides a suitable testing ground for the applications of TPUs to quantum many-body dynamics.

\section{Simulations and results\label{sec:sims_results}}

\begin{figure}
    \centering
    \includegraphics[width=1\linewidth]{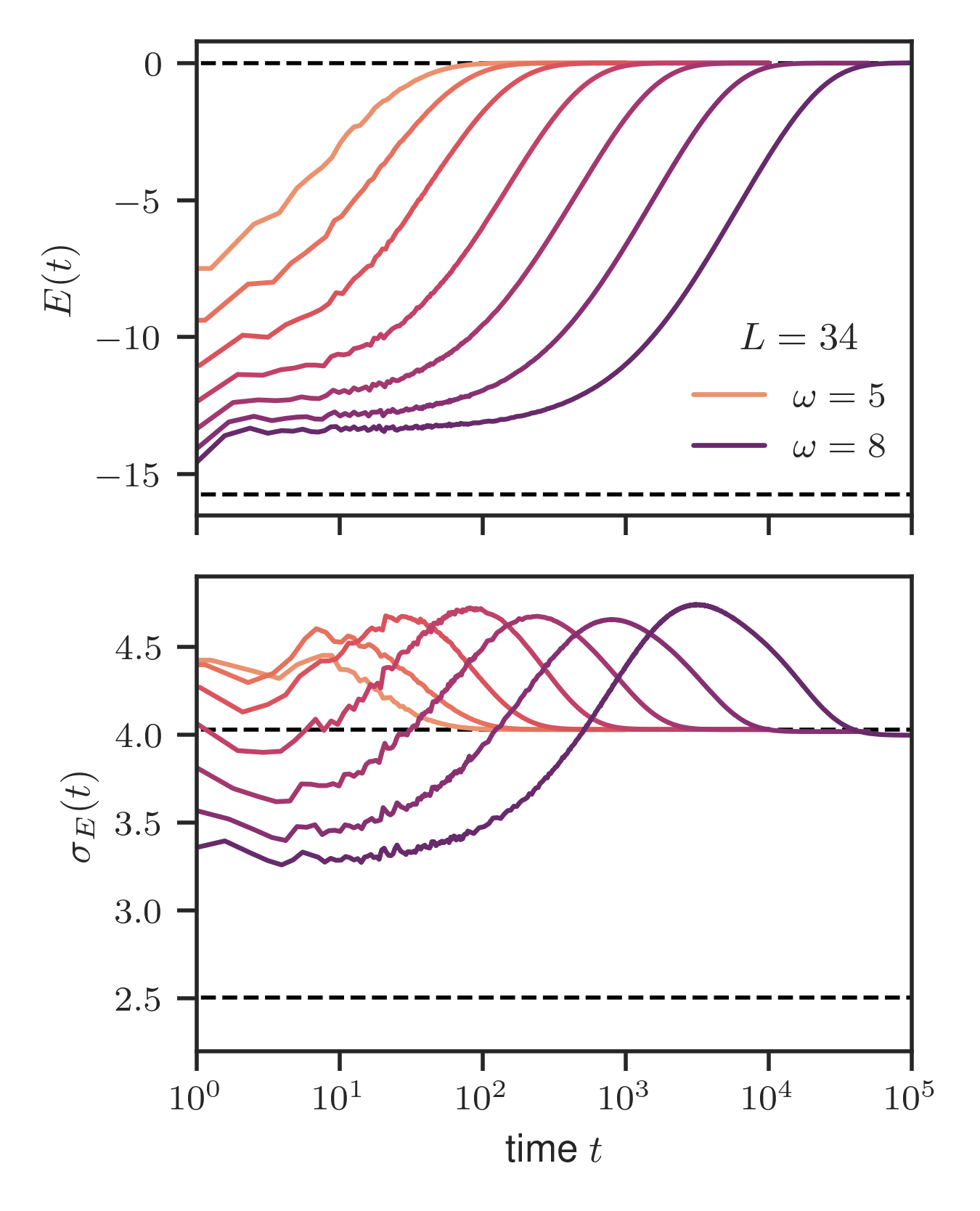}
    \caption{Mean (top) and standard deviation (bottom) of energy, with respect to $\bar{H}$, versus time. Light to dark (left to right) curves correspond to $\omega \in [5,8]$ in steps of $1/2$. All data is for $L = 34$ qubits. Dashed black lines indicate both initial values and infinite-temperature equilibrium values (where the density matrix is $\rho \propto \mathbb{I}$) for comparison. Data was collected after every Floquet period up to $t=10^2$, and then after every tenth period from then on. Note that the conversion from time $t$ to number of gates $g$ is $g = (L-1) t / T$, where $T=2\pi / \omega$, so the rightmost curve corresponds to a total of $4.2\times10^6$ two-qubit gates.}
    \label{fig:energy_mean_and_std}
\end{figure}

We now present numerical evidence for the predicted exponential-in-frequency thermalization times characteristic of Floquet prethermalization. For all of the simulations in this paper, we initialize the system in the ground state of $H_o$. This is a product state of local two-qubit ground states (gs):
\begin{eqnarray}
    |\psi(0)\rangle = \bigotimes_{k\ \text{odd}} |\text{gs}(H_k)\rangle. \label{eq:init_state}
\end{eqnarray}
Since the $H_k$ are uncorrelated and traceless in our model, the energy of this initial state with respect to the full time-averaged Hamiltonian is close to the sum of local ground state energies of the odd terms. We chose this initial state because it is simple and is at a finite energy density corresponding to starting with a finite temperature initial state with respect to $\bar{H}$. 

As accessible indicators of the slow approach to thermal equilibrium and the existence of a quasiconserved $\tilde{H}_F$, we track both the mean and standard deviation of energy with respect to $\bar{H}$: 
\begin{eqnarray}
    E(t) &=& \langle \psi(t)| \bar{H} | \psi(t) \rangle\\
    \sigma_E(t) &=& \sqrt{\langle \psi(t)| \bar{H}^2 | \psi(t)\rangle - E(t)^2}.    
\end{eqnarray}
In Fig.~\ref{fig:energy_mean_and_std} we show $E$ and $\sigma_E$ as a function of time $t$ out to $t = 10^5$ for a system of $L=34$ qubits and for several evenly-spaced values of $\omega \in [5,8]$.
By eye we can see that as $\omega$ increases uniformly, the thermalization timescale increases uniformly on a logarithmic scale, implying that the thermalization rate scales like $\Gamma(\omega) = \exp(a\omega + b)$ within the observable window of frequencies. 

\begin{figure}
    \centering
    \includegraphics[width=1\linewidth]{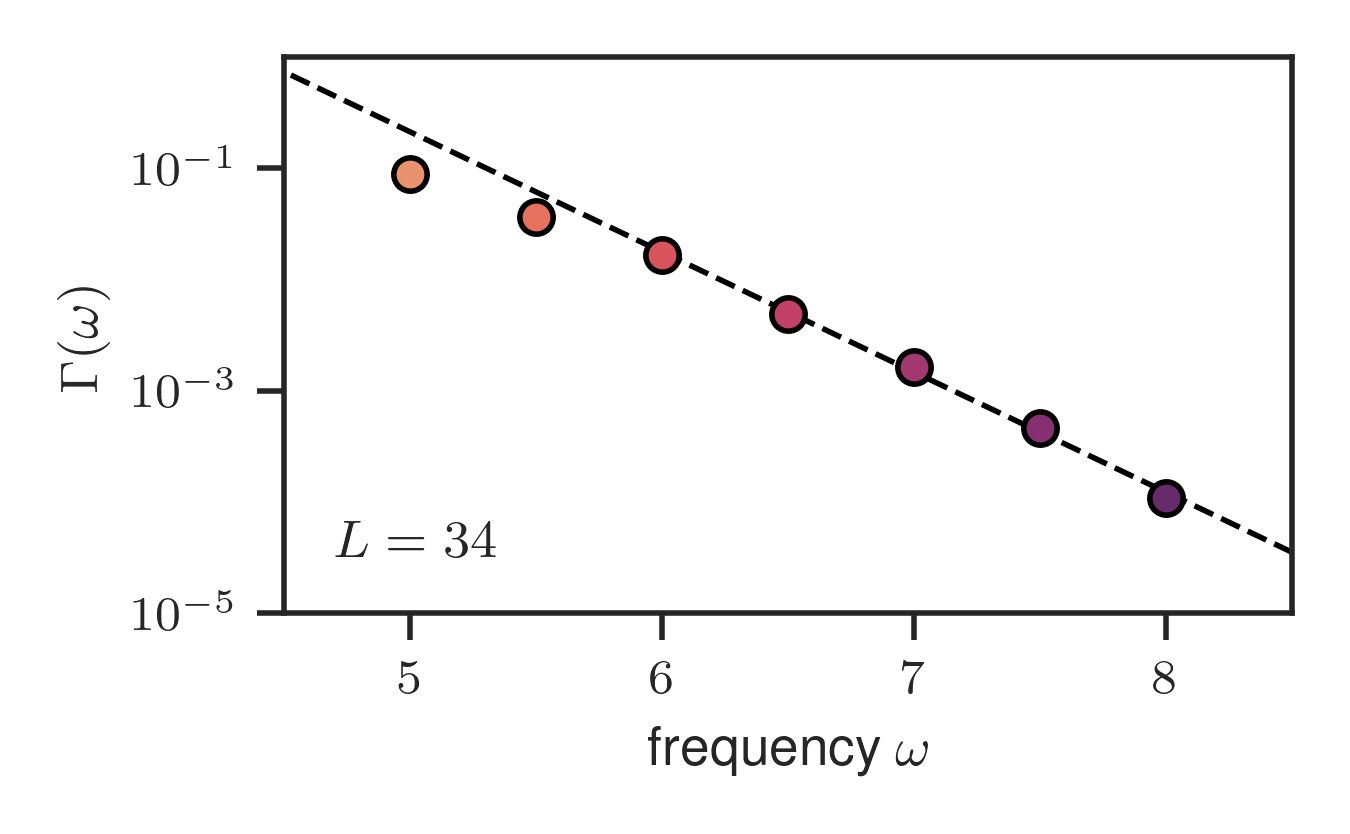}
    \caption{Thermalization rate versus frequency. The marker colors match the corresponding curves in Fig.~\ref{fig:energy_mean_and_std}, and are a redundant encoding of $\omega$, indicated on the horizontal axis. The dashed line is a fit excluding the two leftmost points and results in $\Gamma = \exp(a \omega + b)$ with $a = -2.48(7)$ and $b = 10.9(5)$. Thermalization rates are extracted as described in the main text.}
    \label{fig:thermalization_rate}
\end{figure}

In order to make the frequency dependence quantitative, we extract thermalization rates from the curves in the top panel of Fig.~\ref{fig:energy_mean_and_std}: For each curve (frequency) we determine a decay rate via $\Gamma = 1/(t_2 - t_1)$, where $t_k$ is defined as the time that satisfies $E(t_k) = E(0) \exp(-k)$. Later time intervals would work too, ex: $\Gamma = 1/(t_3 - t_2)$, but we avoid the earliest such interval ($t\in[t_0, t_1]$) because it contains the rapid early-time dynamics of $E(t)$ before it settles into a slowly decaying ``prethermal plateau". In Fig.~\ref{fig:thermalization_rate} we plot these thermalization rates and show that indeed they satisfy an exponential scaling $\Gamma(\omega) = \exp(a \omega + b)$ with $a = -2.48(7)$ and $b = 10.9(5)$. 

Our simulations are in agreement with the theoretical expectation that $\Gamma$ is exponentially suppressed in $\omega$ at large enough frequencies, and with previous experimental~\cite{RubioAbadal-Bloch2020, Peng-Cappellaro2021} and numerical~\cite{Machado-Yao2019, Luitz-Khemani2020, Fleckenstein-Bukov2021} work on other systems. We note here that prior numerical studies of this effect have been performed with system sizes of $L\leq 24$ and times $t\leq 10^4$ (in microscopic units), so using TPUs has allowed us to significantly extend the system sizes and times studied by about ten qubits and a factor of ten in time. This is a rough estimate and not a direct comparison to previous work due to differences in the specific model being simulated and the computational resources consumed (our Floquet circuit model is indeed easier to simulate than non-circuit models, which is one of the reasons for using it). A more direct comparison of computational resources required for simulating our chosen model on CPUs vs. GPUs vs. TPUs is provided in the following section in order to put our work into context.

\begin{figure}
    \centering
    \includegraphics[width=1\linewidth]{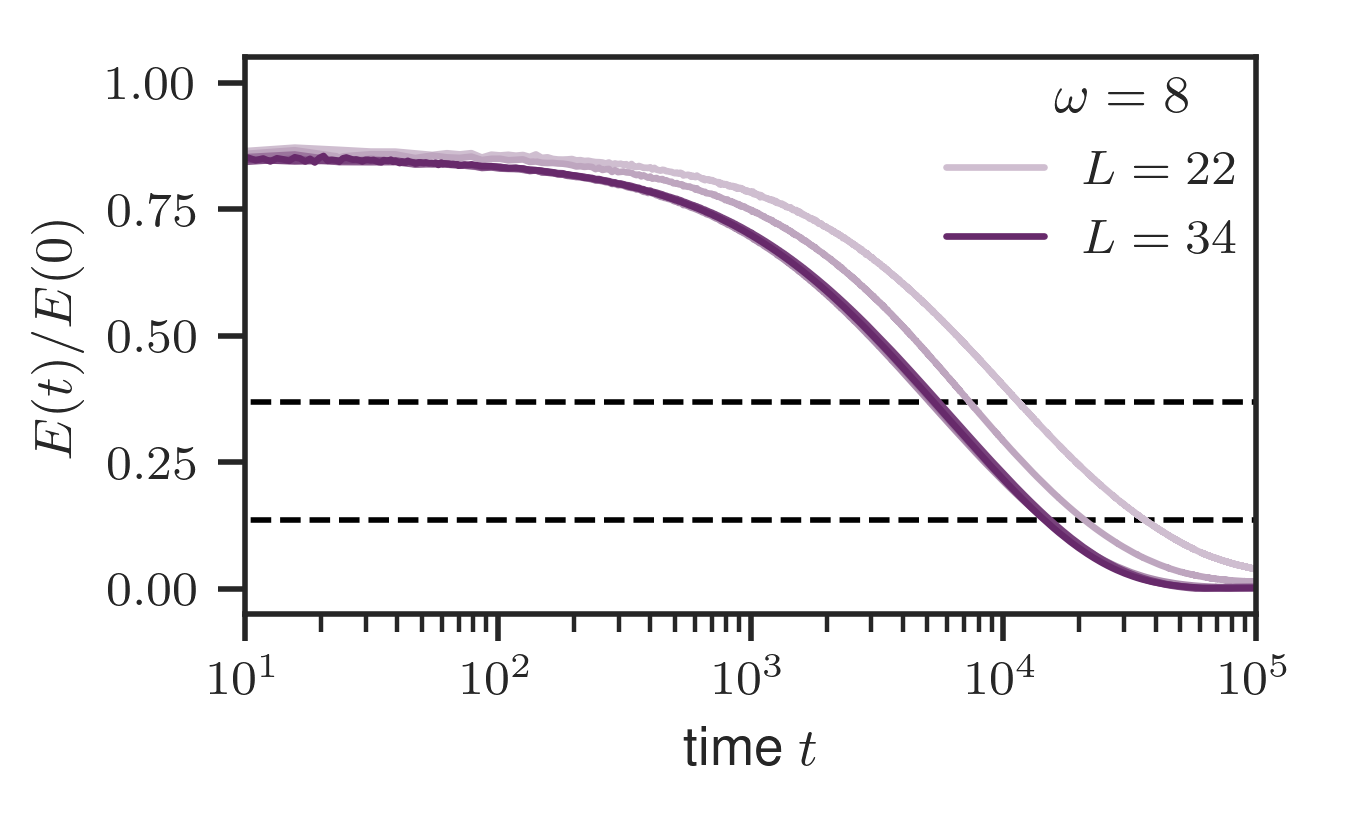}
    \caption{Normalized energy versus time. Light to dark (right to left) curves correspond to $L\in[22,34]$ in steps of $2$. All data is for $\omega = 8$, the largest frequency we consider in Fig.~\ref{fig:energy_mean_and_std}. The dashed horizontal lines are at the thresholds $e^{-1}$ and $e^{-2}$, which are used to extract the thermalization rate $\Gamma$, as described in the main text.}
    \label{fig:convergence}
\end{figure}

In Fig.~\ref{fig:convergence} we demonstrate sufficient convergence in the number of qubits $L$ to extract large-$L$ decay rates. We fix the frequency of the drive to $\omega = 8$, the largest frequency we consider, and plot $E(t)/E(0)$ over time for system sizes $L\in[22,34]$. The convergence of these curves at the larger values of $L$ ensures that the thermalization rates shown in Fig.~\ref{fig:thermalization_rate}, which are for systems with $L=34$, are converged to the $L\to \infty$ result. In Sec.~\ref{sec:cost_accuracy} we demonstrate simulations of up to $38$ qubits for much shorter times, in order to study the computational cost of these simulations.

Some clarifying comments on our presented data are in order. First, the initial energy $E(0)$ is indeed independent of $\omega$, even though it may not appear that way in Fig.~\ref{fig:energy_mean_and_std} due to the logarithmic horizontal scale. This is because all simulations start in the same state (Eq.~\ref{eq:init_state}), and we are using only the zeroth order effective Hamiltonian $\tilde{H}_F^{(0)} = \bar{H}$ to compute energy. The fact that $E(t)$ does vary significantly with $\omega$ but slowly with $t$ at early times is a reflection of the fact that $\bar{H}$ is not a very good approximation to the effective Hamiltonian $\tilde{H}_F$, even in the regime in which the heating time is exponential in $\omega$. Another indicator that $\bar{H}$ is quite different from the optimal effective Hamiltonian is that $\sigma_E$ takes values above the infinite temperature equilibrium value (bottom panel of Fig.~\ref{fig:energy_mean_and_std}). Generally, a thermal state of a Hamiltonian $H$ has a larger energy variance at infinite temperature than at any finite temperature. The fact that we see larger values of $\sigma_E$ is fully consistent with the system being in a quasistatic (with slowly evolving temperature) thermal state of an ideal prethermal Hamiltonian $\tilde{H}_F$ that differs from $\bar{H}$ by corrections that are suppressed in $\omega$, as expected. Finally, the rightmost curve in the bottom panel of Fig.~\ref{fig:energy_mean_and_std}, corresponding to $\omega = 8$, converges to a value that is visibly slightly different from the infinite-temperature value. This is because at $\omega = 8$ the system is just starting to enter the finite-size regime, for $L=34$, where the system has difficulty exchanging energy with such a high-frequency drive. It is interesting to note that this effect is present for $\sigma_E$ but not for $E$, and thus higher moments of the energy distribution seem to be more sensitive to any residual nonthermalizing dynamics.

\section{Computational cost and numerical accuracy \label{sec:cost_accuracy}}
\begin{figure}
    \centering
    \includegraphics[width=1\linewidth]{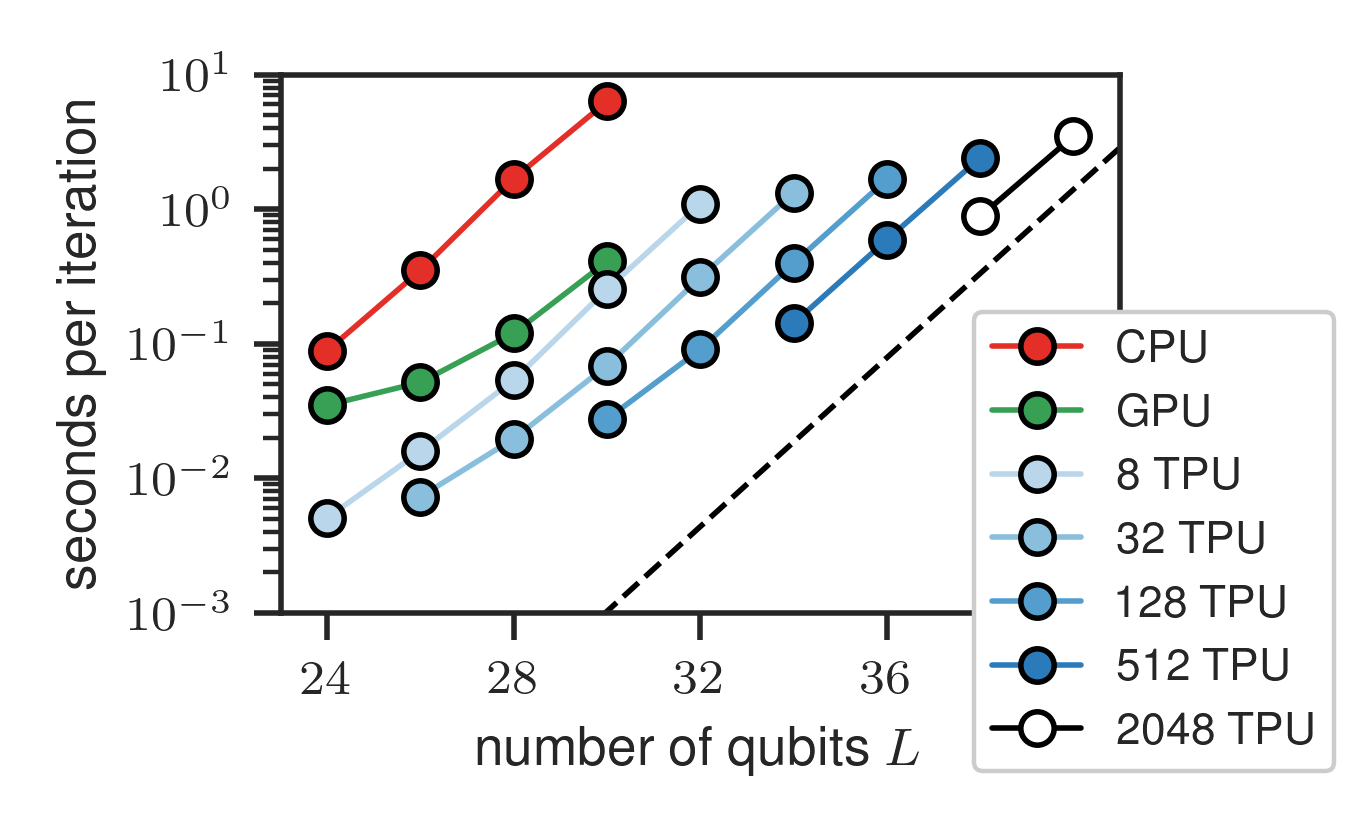}
    \caption{Wall-clock runtime for applying the Floquet unitary circuit (two layers of gates) to an $L$-qubit wavefunction. The red data in the top left are reference values obtained by simulating the circuit on a standard 8-CPU-core workstation using \texttt{qsimcirq}~\cite{qsim, cirq}. The green data are \texttt{qsimcirq} simulations run on one NVIDIA A100 GPU. The blue data, from left to right, corresponds to 8, 32, 128, and 512-TPU-core clusters. The slope of the dashed line indicates the expected asymptotic (large $L$) scaling of the computational cost, $\propto L 2^L$. The unfilled markers in the top right are extrapolated from the top two points in each series, and represent our maximum capacity simulations on 2048 TPU cores. All simulations were performed with single-precision arithmetic. All TPU calculations in this paper were performed on the third generation of TPUs, ``TPU v3".
    }
    \label{fig:wall_times}
\end{figure}

The workhorse of our simulation method is the application of the two-layer Floquet unitary circuit to the wavefunction, $|\psi(nT)\rangle = U_F |\psi([n-1]T)\rangle$, using TPUs to accelerate and scale the computation. Applying the Hamiltonian is also crucial, for computing $E$ and $\sigma_E$, but relies on the same technology. We therefore provide a summary of the wall-clock time it takes to evolve an $L$-qubit wavefunction by one cycle of the circuit in Fig.~\ref{fig:wall_times}. The timing is done by simulating 100 successive periods and dividing the total execution time by 100. Since we are not simulating out to long times, as in Sec.~\ref{sec:sims_results}, we provide results for simulations of up to $L=38$ qubits done on clusters with up to 512 TPU cores, and extrapolate to our maximum capacity of $L=40$ on 2048 TPU cores. We also perform reference CPU and GPU simulations on up to $30$ qubits. Of course, there is no single correct metric for comparing CPU to GPU to TPU simulations, but we believe our reference CPU and GPU simulations are more performant than what most researchers use in the quantum dynamics and thermalization community~\cite{quspin,Lezama-Bardarson2019,Machado-Yao2019}, and our purpose here is to demonstrate the application of TPUs in that setting.

In Fig.~\ref{fig:wall_times} we see that, for the same number of qubits, simulations on an 8-TPU-core machine are about $25\times$ faster than reference simulations using the \texttt{qsimcirq} package on a standard 8-CPU-core workstation (Intel Xeon @ 2.3 GHz with 64 GB of RAM), and faster than or comparable to the \texttt{qsimcirq} simulations performed on one A100 GPU. For example, for $L=30$ our reference CPU (GPU) calculation took 6.3 seconds (0.41 seconds) per cycle, whereas the 8-TPU-core machine took 0.25 seconds. A significant further speedup is shown by distributing the computation over 128 TPU cores, which results in a wall-clock time that is $230\times$ and $15\times$ shorter than the reference CPU and GPU calculations, respectively. 

Beyond this speedup, the easy distribution of the computation over many TPUs (see Sec.~\ref{sec:algorithms}) allows for a higher total memory (up to 32 TB) and thus larger system sizes (up to $L=40$). As shown in Fig.~\ref{fig:wall_times}, the cost of distributing the calculation over many TPUs is quite tolerable, and therefore our method is highly scalable. For example, the $L = 32$ simulation was done on 8, 32, and 128-TPU-core clusters, and the speedup factors in going from 8 to 32, and 32 to 128 TPU cores were $3.5\times$ and $3.4\times$, compared to the factor of $4$ we would expect if inter-TPU communication was negligible. Another way to see this feature is by looking at the top point in each series of TPU data. By scaling up the number of TPU cores by a factor of $2^6$ we were able to scale the computation from 32 to 38 qubits (Hilbert space grows by a factor of $2^6$ here) while only increasing the wall-clock time by a factor of $2.2\times$. In the ideal case where inter-TPU communication is negligible, this factor would be approximately $38/32\approx 1.2$, because the total number of gates increases proportionally with the number of qubits. 

Since for these simulations an 8-TPU-core machine is comparable to an A100 GPU, we expect that a full TPU pod of 2048 cores is comparable to a configuration of hundreds of GPUs (currently the largest multi-GPU configuration available on Google Cloud has sixteen A100 GPUs). Thus large TPU clusters seem to have strengths of being more practically available and more energy and cost efficient than comparable GPU-based machines. We also note that the latest TPU v4 pod has twice the matrix-multiplication throughput per chip, and four times as many chips (connected via state-of-the-art interconnects), as the v3 used in this work. Thus TPUs are a state-of-the-art platform for the kind of quantum simulations we study in this work.

\begin{figure}
    \centering
    \includegraphics{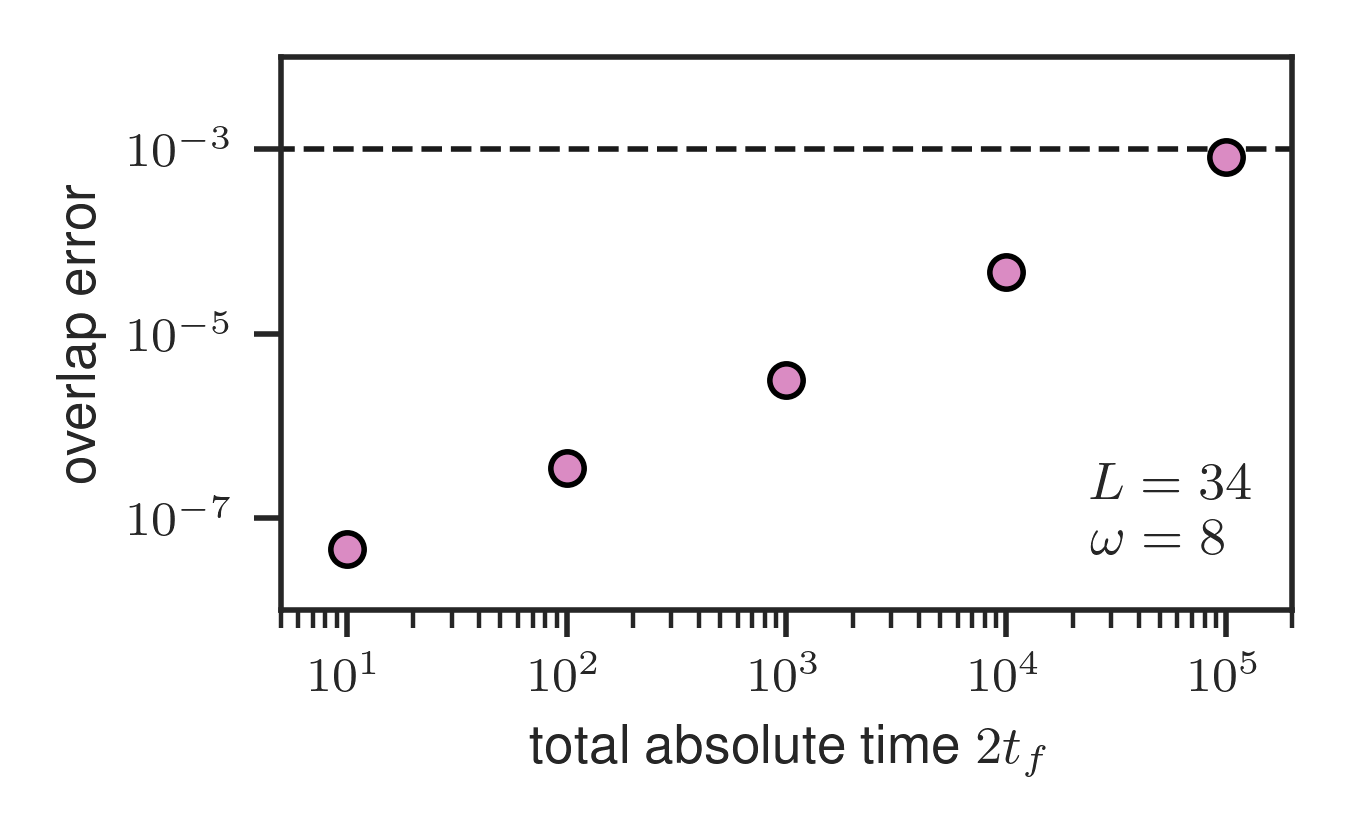}
    \caption{Accumulation of error in time. The error on the wavefunction is quantified via the overlap of initial and final states in an ``echo" procedure, where the wavefunction is evolved out to time $t_f$, and then the evolution is reversed using the time-reversed Floquet unitary $U_F^\dagger$ in order to get back to time $t=0$. The quantity we show is $|1 - \langle \psi(0)|\psi'(0) \rangle|$, where $|\psi'(0)\rangle$ is the state obtained from evolving $|\psi(0)\rangle$ out to time $t_f$ and back. All simulations are run at $\omega=8$ and $L=34$. The dashed black line is at the 0.1\% error level, for reference.}
    \label{fig:relative_errors}
\end{figure}

Beyond the cost of TPU simulations, we must also verify their accuracy, especially for the kind of long-time simulations presented in Sec.~\ref{sec:sims_results} because TPUs were designed for fast, low-precision matrix multiplication, among other things~\cite{TPUinfo}. While double-precision (64-bit) arithmetic can be emulated at a cost (about $11\times$ slower than single-precision), the simulations presented in this work were all performed in single-precision (32-bit). Because of this, numerical errors are larger than in double-precision simulations, which are the standard in quantum dynamics research. While the required precision is task dependent, and there are quantum simulations that require higher precision than single or even double, single-precision arithmetic is sufficient for many calculations and comes with the advantage of being less costly. In order to study the accumulation of error, and validate our earlier results, we perform an ``echo" procedure: We first time evolve the usual initial state $|\psi(0)\rangle$ (Eq.~\ref{eq:init_state}) out to time $t=t_f$, then reverse the time evolution back to time $t=0$, using the time-reversed Floquet unitary $U_F^\dagger$, in order to obtain $|\psi'(0)\rangle$. Since the forward and backward evolutions are carried out with a finite numerical precision, $|\psi(0)\rangle$ and $|\psi'(0)\rangle$ are not exactly the same. We quantify the difference via the ``overlap error" $|1 - \langle \psi(0) | \psi'(0)\rangle|$. We have checked on smaller scale CPU simulations that this is a good way to estimate $|1-\langle\psi(2t_f)|\psi'(2t_f)\rangle|$, where here the reference state $|\psi(2t_f)\rangle$ is given by double-precision simulation and $|\psi'(2t_f)\rangle$ by single-precision. Since the overlap (fidelity) is a good measure of the proximity of two wavefunctions, and the wavefunction contains all of the information about a quantum system, this is a robust way of tracking the accumulation of numerical error over time. In Fig.~\ref{fig:relative_errors} we show the overlap error resulting from this echo procedure as a function of the total simulated time, $2t_f$, at $L=34$ and $\omega=8$. From Fig.~\ref{fig:relative_errors} we see that indeed our simulations are accurate up to an overlap error of about $10^{-7}$ at early times (as expected with 32-bit precision), and at the latest time $t=10^5$ the overlap error reaches $10^{-3}$. We also find that the \textit{relative} error in $E(0)$ and $\sigma(0)$, when comparing $|\psi'(0)\rangle$ to $|\psi(0)\rangle$, are comparable to the overlap error for these simulations. This serves to validate our earlier data and confirm the applicability of single-precision TPU simulations for studying Floquet prethermalization and other similar problems. We note that some problems do require more than single-precision arithmetic, and in that case TPUs may offer an advantage of scalability even though emulating double-precision is costly, or TPUs may not be appropriate in their current form for those problems.

\section{Summary and discussion \label{sec:summary}}
The main goal of this work was to demonstrate a new approach to the classical simulation of quantum many-body dynamics that leverages Google's TPUs, and associated software, to accelerate and scale the computation.

We first gave an overview of the principles of TPU algorithms for quantum simulation, then briefly reviewed the phenomenon of Floquet prethermalization and exponentially slow heating in periodically driven quantum spin chains, and proposed a minimal Floquet circuit model that exemplifies this phenomenon. We applied our methods to demonstrate this exponential scaling over longer times and on larger systems than had previously been accessible: In this study we computed nontrivial dynamics, accurately, on long timescales (over $10^5$ Floquet cycles) for systems of $34$ qubits using a 128-TPU-core cluster, and showed that by distributing the computation over 2048 TPU cores, systems with up to $40$ qubits are accessible. We also demonstrated a $230\times$ ($15\times$) speedup in wall-clock runtime when comparing a distributed TPU simulation to a reference simulation performed on a CPU (GPU). This was intended to provide familiar points of reference, and not as proof that TPUs are the most performant hardware. For that, an extensive comparison to multi-GPU configurations would be necessary. However, as mentioned earlier, we expect that a full TPU pod is comparable to a machine with hundreds of top-of-the-line GPUs, and such configurations are not readily available. We emphasize that in this problem, entanglement quickly becomes nearly maximal, and so keeping track of all $2^L$ amplitudes of the wavefunction is necessary.

The problem of studying Floquet prethermalization is a good example of the challenging computational tasks that arise in quantum many-body dynamics research, and it provides a suitable testing ground for demonstrating the application of TPUs in this field. TPU simulations widen the finite-size and finite-time windows that numerical simulations can access, thereby allowing for higher-quality data and more thorough studies of dynamical phenomena and phase transitions in quantum many-body systems. In the future, TPU simulations of the full wavefunction, as we have done here, could be useful for many topical studies of quantum dynamics, wherein our understanding of many fundamental issues is limited by finite system size and time studies~\cite{Suntajs-Vidmar2020,KieferEmmanouilidis-Sirker2021,SelsPolkovnikov2020, Abanin-Vasseur2021,Sierant-Zakrzewski2020,Panda-Znidaric2020,Luitz-BarLev2020,Morningstar-Huse2021,sierant2021observe}.

A natural extension of this work is to employ TPU methods for performing state-of-the-art classical simulations for the purposes of benchmarking simulations done on quantum hardware, and in assessing quantum advantage. For this, entanglement may not be near maximal as it is for the Floquet prethermalization problem, so it would be promising to also develop TPU-accelerated versions of tensor network methods. The development of tensor network methods may also benefit from having larger scale exact simulations, like the ones performed in this work, to compare to.

\begin{acknowledgments}
This research was supported with Cloud TPUs from Google's TPU Research Cloud (TRC).
Sandbox is a team within the Alphabet family of companies, which includes Google, Verily, Waymo,
X, and others. The GPU simulations presented in this article were performed on computational resources managed and supported by Princeton Research Computing. GV is a CIFAR fellow in the Quantum Information Science Program and a Distinguished Visiting Research Chair at Perimeter Institute. Research at Perimeter Institute is supported by the Government of Canada through the Department of Innovation, Science and Economic Development and by the Province of Ontario through the Ministry of Research, Innovation and Science. V.K. acknowledges support from the Sloan Foundation through a Sloan Research Fellowship, the Packard Foundation through a Packard Fellowship, and by the US Department of Energy, Office of Science, Basic Energy Sciences, under Early Career Award No. DE-SC0021111.

\end{acknowledgments}

\bibliography{main}

\end{document}